\documentclass[runningheads,11pt]{llncs}
\usepackage[T1]{fontenc}
\usepackage{graphicx}
\usepackage{color}
\usepackage{color, colortbl}
\definecolor{Gray}{gray}{0.9}
\usepackage{amssymb}

\begin{document}
\title{Private Information Retrieval Schemes Using Cyclic Codes\thanks{This work was supported in part by Grant PGC2018-096446-B-C21 funded by MCIN/AEI/10.13039/501100011033 and by ``ERDF A way of making Europe'', by Grant RYC-2016-20208 funded by MCIN/AEI/10.13039/501100011033 and by ``ESF Investing in your future'', and by Grant CONTPR-2019-385 funded by Universidad de Valladolid and Banco Santander.}}

%
%\titlerunning{Abbreviated paper title}
% If the paper title is too long for the running head, you can set
% an abbreviated paper title here
%
\author{\c{S}eyma Bodur%\inst{1}%\orcidID{0000-0002-6171-7975} 
\and
Edgar Mart\'inez-Moro%\inst{1}%\orcidID{0000-0003-1243-2049} 
\and
 Diego Ruano%\inst{1}%\orcidID{0000-0001-7304-0087}
 }
\authorrunning{\c{S}. Bodur et al.}
% First names are abbreviated in the running head.
% If there are more than two authors, 'et al.' is used.
%
\institute{IMUVa-Mathematics Research Institute,\\ Universidad de Valladolid, Spain \\
%\url{http://www.imuva.uva.es}\\
\email{\{seyma.bodur,edgar.martinez,diego.ruano\}@uva.es}
}

\maketitle              % typeset the header of the contribution
\begin{abstract}
A Private Information Retrieval (PIR) scheme  allows users to retrieve data from a database without disclosing to the server information about the identity of the data retrieved. A coded storage in a distributed storage system with colluding servers is considered in this work, namely the approach in \cite{RM} which considers a storage and retrieval code with a transitive group and provides binary PIR schemes with the highest possible rate. Reed-Muller codes were considered in \cite{RM}. In this work we consider cyclic codes and we show that binary PIR schemes using cyclic codes provide a larger constellation of PIR parameters and they may outperform the ones coming from Reed-Muller codes in some cases.
\end{abstract}
\keywords{Private information retrieval  \and Cyclic codes \and Reed-Muller codes.}
\section{Introduction}
Many protocols protect the user and server from third parties while accessing the data. Nevertheless, no security measure protects the user from the server. As a result of this demand, Private Information Retrieval (PIR) protocols emerged \cite{origin}. They allow users to retrieve data from a database without disclosing to the server information about the identity of the data retrieved. We consider in this work that data is stored in a Distributed Storage System (DSS), since, if data is stored in a single database, one can only guarantee information-theoretic privacy by downloading the full database, which has a high communication cost.

Shah et al.~\cite{five} have shown that privacy is guaranteed when one bit more than the requested file size is downloaded, but it requires many servers. In case of non-response or a fail from some servers, the PIR scheme should allow servers to communicate with each other. Hence, it is natural to assume that the servers may collude, that is, they may inform each other of their input from the user. A scheme that addresses the situation where any $t$ servers may collude is called a $t$-private information retrieval scheme and it was considered in \cite{siam,Tajadine2018}. This approach, that we consider in this work, uses Coding Theory and the security and performance depend on the parameters of linear codes and their star products (also called Shur products).

%\textcolor{red}{One reviwer says: In [4], a PIR scheme with colluding server capacity for GeneralizedReed Solomon (GRS) codes is given.". This is not true, the capacity was never settled until now. Please correct.}

The PIR maximum possible rate was examined without collusion in \cite{SunCapacity1} and with collusion in \cite{SunCapacity2}. The PIR capacity obtained without colluding servers and using a Maximum Distance Separable (MDS) code was given in \cite{twelve}. In \cite{siam}, a PIR scheme with colluding servers %capacity 
for Generalized Reed Solomon (GRS) codes is given. Their PIR scheme rate is based on the minimum distance of a star product of the storage code and the retrieval code.

The use of a GRS, or an MDS code, requires working over a big base field.  In order to address this issue, since binary base fields are desirable for practical implementations, \cite{RM} provided a PIR scheme that is based on binary Reed-Muller (RM) codes. They observed that the scheme reaches the highest possible rate if the codes used to define the PIR scheme have a transitive automorphism group, which is the case of RM codes.

In this work we propose to use cyclic codes to construct PIR schemes in the same fashion as \cite{RM}. Cyclic codes have also a transitive automorphism group and they can be defined over a binary (or small) finite field as well. Moreover, the star product of two cyclic codes is a cyclic code and its parameters can be computed \cite{Cascudo}. Namely, the star product of two cyclic codes is given by
the sum of their generating sets and we can compute its dimension and estimate its minimum distance considering cyclotomic cosets.

The main contributions of this work are given in Section~\ref{sec:principal}. Our aim is to optimize the number of databases that can collude without disclosing to the server information about the identity of the data retrieved. In order to show the goodness of cyclic codes for PIR schemes, we first provide pairs of cyclic codes $C$ and $D$, the storage code and retrieval code, such that the parameters of $C$, $D$, $D^\bot$, $C\star D$ and $(C\star D)^\bot$ are -at the same time- optimal or the best known. As we will recall in Section~\ref{sec:PIR}, their parameters determine the performance of the PIR scheme defined by $C$ and $D$. Since a punctured RM code is a cyclic code, we may obtain PIR schemes using punctured RM codes by using cyclic codes. Moreover, we show that by using cyclic codes we obtain a larger constellation of possible parameters of binary PIR schemes. The construction of PIR schemes and the computations of their parameters follow from a detailed analysis of cyclotomic cosets. Then we focus on the privacy and on the rate of a PIR scheme since the upload cost in a PIR scheme can be neglected \cite{twelve}. More concretely, in case that the storage code $C$ has dimension 2, we obtain binary PIR schemes that greatly outperform the ones  obtained using RM codes, more concretely they protect against a more significant number of colluding servers. Finally, we compare  our schemes with shortened RM codes and we show that in this case the PIR schemes using cyclic codes outperform them as well, namely they offer more privacy for a fixed rate.

%\textcolor{red}{One reviewer claims: The only actual advantage I could  come up with would be a reduced cost in generating the query vectors since a smaller dimension of the code D means less randomness needs to be generated by the user. But this is not mentioned in the paper.}

%\textcolor{red}{I do not understand very well what it says but we should look at it} Done, see the "conclusions" section

\section{General Private Information Scheme}\label{sec:PIR}
This section reviews some basic definitions of linear codes and briefly recalls PIR schemes (see \cite{RM,siam,Tajadine2018} for further details). We denote by $\mathbb F_q$  the finite field with $q$ elements. A linear code $C$ is a linear subspace of $\mathbb{F}_q^n$. We denote its parameters by $[n,k,d]$.

%\begin{definition}%\end{definition}

%The following definition is essential for characterizing the security of a PIR scheme.

\begin{definition} Given two linear codes $C$ and $D$ of length $n$ over $\mathbb{F}_q$, we define their star product (or Shur product) $C\star D$ as the linear code in $\mathbb{F}_{{q}}^n$ spanned  by the set $\{  c\star d \; | \; c\in C, d\in D  \}$, where $\star$ denotes the component-wise product $c\star d = (c_1 d_1, \ldots , c_n d_n)$.
\end{definition}
\subsection{PIR Schemes}
A PIR scheme consists of three stages; Data Storage, File Request and Response Process. In the Data Storage Process, files are uploaded to a  DSS. In the File Request, users decide the file they want to retrieve, called the desired file, and according to it, they select  queries  that are sent to the servers. In the final Response Process, servers `operate' the files with the queries generating a matrix of responses that are sent back to the user. It should be noted that the servers do not have any information about the file requested by the user.

%\textcolor{red}{One reviewer claims: (5) Section II.This section cannot be fully understood without specifying how storage is done, e.g., "servers multiply the files corresponding to the random codeword received and thensend this matrix back to the user." It is unclear what matrix are we referring to without discussing the storage code. Please consider re-ordering the sections. }

\subsubsection{Data Storage Process}

We have $r$-files, each file has $\rho$-rows, $k$-columns, and the elements of the files are in $\mathbb{F}_{{q}}$. Since the number of files is $r$, the total file can be understood as $r \rho \times k$ matrix denoted by $A$, and each file is denoted by \textbf{${\mbox{\boldmath $a$}}^i$}, where $i \in \{1,\dots, r\}$.

The files are stored in a DSS. In order to upload the files into the servers, files are encoded by a $k$-dimensional storage code $C \subseteq \mathbb{F}_{{q}}^n$ with parameters $[n,k,d]$. Concerning encoding, we multiply the matrix $A$, which covers all files, by $G_C$ the generator matrix of the linear code $C$ and we obtain the matrix $Y:= A \cdot G_C$. Since $A$ is a $\rho r \times k$  matrix,  $Y$  has $\rho r$ rows and $n$ columns.

\subsubsection{Request and Response Process}
Let's assume that the user wants to retrieve the file ${\mbox{\boldmath $a$}}^i$. Then the user chooses a random query $Q^i$ and sends this query to the servers. Each server computes the inner product of $Q^i_j$ and  $Y_j$, where $j$ is the server index, i.e, $j^{th}$-server computes $\langle Q^i_j, Y_j \rangle$. Then, servers send back the response vectors to the user.

The file is divided into parts, and a part is obtained in each round. With the final round, all parts of the file are completed. All the parts of the file from several servers are gathered to get the whole file. Therefore, the PIR rate is defined as the ratio of the information obtained during the process to the downloaded information.

If $t$-colluding servers communicate with each other and they cannot access any information about the desired file, it is said that the PIR scheme is resistant to $t$-colluding servers.
The following theorem is the key for finding the number of colluding servers and the system's PIR rate.

\begin{theorem}[\cite{RM}]\label{theo:1}
If the automorphism groups of $C$ and $C\star D$ are transitive on the set $\{1,\dots,n\}$, then there exists a PIR scheme with rate $\frac{dim(C\star D)^{\bot}}{n}$ that resists a $(d_{D^{\perp}}-1)$--collusion attack. That is, the privacy is $t=d_{D^{\perp}}-1$.\end{theorem}

We will compare Reed-Muller codes, considered in \cite{RM}, and cyclic codes in Section~\ref{sec:principal}. For this reason, the next section gives a brief exposition of RM codes.

\section{Reed-Muller Codes}
The binary $r^{th}$ order Reed-Muller Code, denoted by $RM(r,m)$, is defined to be
\begin{equation}
RM(r,m)= \left\{ ev(f) :  f\in \mathbb{F}_2[x_1,...,x_m],\; deg(f) \leq r\right\},
\end{equation}
where $ev(f)$ is the evaluation of $f$ at all points in  $\mathbb{F}_2^m$.

\begin{remark}  $RM(r,m)$ is a linear code of length $n=2^m$, dimension $k=\sum_{i=0}^r {m \choose i}$, and minimum distance $2^{m-r}$ \cite{eight}. One has that $RM(r_1,m)\; \star \; RM(r_2,m) = RM(r_1+r_2, m)$, where $r_1+r_2 \leq m$.

\end{remark}

Since if we shorten or puncture  RM codes at the position evaluating at $\mathbf 0$ we obtain cyclic codes \cite{four}, the following two definitions will be helpful in Section~\ref{sec:principal}.
\begin{definition} The shortened code at the position evaluating at $\mathbf 0$ of a binary Reed-Muller code is denoted by the linear $[2^{m}-1, k-1, 2^{m-r}]$ code $C_{\bullet}$. The punctured code at the position evaluating at $\mathbf 0$ of a binary Reed-Muller code is denoted by the linear $[2^{m}-1, k, 2^{m-r}-1]$ code $C^{\bullet}$.
\end{definition}

%Due to  our PIR scheme is based on cyclic codes, Section~\ref{sec:IV} presents some preliminaries on that class of codes.

\section{Cyclic Codes } \label{sec:IV}
In this section, we will be concerned with basic cyclic code definitions and compute the star product of two cyclic codes.

%\textcolor{red}{WE HAVE NOW ROOM TO EXPLAIN CYCLIC CODES A BIT MORE. IN PARTICULAR WE SHOULD EXPLAIN THE ROLE OF CYCLOTOMIC COSETS IN CYCLIC CODES.}
%\textcolor{red}{$n$ and $q$ should be relatively prime, one reviewer recalls}

\begin{definition}A $[n,k]$ linear code $C$ is said to be cyclic if every cyclic shift of a codeword $c=(c_0,c_1,\dots, c_{n-1})\in C$ is also codeword in $C$, that is $c=(c_{n-1},c_0,\dots, c_{n-2})\in C$.\end{definition}

\begin{theorem}[\cite{six}] A linear code $C$ is a cyclic code if and only if $C$ is isomorphic, as a $\mathbb F_q$-linear space, to  an ideal in the ring $R_n = \mathbb{F}_q[x]/ \langle x^n-1\rangle$.\end{theorem}
In order to have a semisimple algebra, i.e, non-repeated roots for the polynomial $ x^n-1$ and a 1-1 correspondence between it factors and its roots, from now on we will require that $\mathrm{gcd}(n,q)=1$.

\begin{definition} Let $C$ be a cyclic code in $R_n$. We call $g(x)$ the generator polynomial of $C$ if there exists a unique monic polynomial $g(x)$ such that $C=\langle g(x)\rangle$. Clearly, $g(x)$ is a divisor of $x^n-1$ in $\mathbb{F}_q[x]$.\end{definition}

\begin{definition}  A set $J\subseteq \{ 0,\ldots, n-1\}$ is said to be the defining set of $C=\langle g(x)\rangle$ if $J= \{ \; j\in \mathbb{Z}/n\mathbb{Z} \mid g(\alpha^j)=0 \; \} $  and a set $I$ is called the generating set of $C=\langle g(x)\rangle$  if  $I= \{ \; j\in \mathbb{Z}/n\mathbb{Z} \mid  g(\alpha^j)\neq 0 \; \} $, where $\alpha$ is a primitive element of $\mathbb{F}_q$.\end{definition}

\begin{remark} Let $g(x)$ be a generator polynomial of cyclic code $C$ then
$
g(x)= \prod_{j\in J}(x-\alpha^j)$ and $g(x) = \frac{x^n -1}{\prod_{i\in I}(x-\alpha^i)}.
$
Furthermore, one has that\\$dim(C)=n-|J|=|I|$.\end{remark}
\begin{remark} \label{rem:1} Assume that $J$ is a defining set of the code $C$, then the  generator polynomial of $C^{\bot}$ is $h(x)=\prod_{i\in -I}(x-\alpha^i)$, where $-I$ is the set of additive inverses in $\mathbb{Z}/n\mathbb{Z}$ of the elements in  $I$.
\end{remark}
\begin{definition} The cyclotomic coset containing $s$, denoted by $U_s$, is defined to be the set $\{s, sq,\dots, sq^i\}\pmod{n}$ where $i$ is the smallest integer such that $q^i\equiv 1 \pmod{n}$.\end{definition}

%Thanks to cyclotomic cosets, we can compute the cyclic code's dimension and estimate its minimum distance.

% To obtain a PIR scheme from cyclic codes,
We have the following result about the star product of cyclic codes.

\begin{theorem} Let $I_1$ and $I_2$ be the  generating sets of the cyclic codes $C$ and $D$, respectively. % Thus, $C$ is generated by $g_1(x) = \frac{x^n -1}{\prod_{i_1\in I_1}(x-\alpha^{i_1})}$ and $D$ is generated by  $g_2(x) := \frac{x^n -1}{\prod_{i_2\in I_2}(x-\alpha^{i_2})}$.
The star product of $C\star D$ is generated by
\begin{equation}
g_{C\star D} = \frac{x^n -1}{\prod_{j\in I_1+I_2}(x-\alpha^{j})},
\end{equation}
where $+$ denotes the Minkowski sum on sets, that is,$$I_1+I_2 := \{ i_1+i_2 \mid i_1\in I_1,\;  i_2 \in I_2\}.$$\end{theorem}

\begin{proof} We will  follow  a similar way to  \cite[Theorem III.3]{Cascudo}, which proves this result for $C\star C$.
It is well known from \cite{Bierbrauer} that a cyclic code can be defined as follows, consider $\mathbb K$ the extension field of $\mathbb F_q$ such that $x^n-1$ splits in linear factors in $\mathbb K[x]$. For a set $M\subseteq \{1, \ldots, n-1\}$ let $\mathcal B(M)$ be the $\mathbb K$-vector space
$$ \mathcal B(M)=\left\{(f(\alpha^0), f(\alpha^1),\ldots, f(\alpha^{n-1}))\mid f=\sum _{i\in M} f_i x^i\in\mathbb K[x]\right\}.$$
For a cyclic code $C$ with defining set $I$, as a byproduct of Delsarte's theorem, one has that $C$ is equal to the subfield subcode $ \mathcal B(-I)\vert_{\mathbb F_q^n}=B(-I)\cap{\mathbb F_q^n}$ (see \cite[Lemma 5]{Cascudo}).  
Now note that the vector space obtained by the extension of scalars of $ C$, denoted by   $\mathbb K\otimes C$, is a $\mathbb K$-cyclic code with the  same dimension as $\mathcal B(-I)$ (given by $|I|$) and, henceforth  $(\mathbb K\otimes C)=\mathcal B(-I)$. Note that the extension by scalars commutes with the star product (see \cite[Lemma 2.23]{Randriambololona}) thus it is clear that 
$C\star D=(\mathbb K\otimes (C\star D))\vert_{\mathbb F_q^n}=(\mathbb K\otimes C\star \mathbb K\otimes   D)\vert_{\mathbb F_q^n}=\left(\mathcal B(-(I_1))\star \mathcal B(-(I_2))\right)\vert_{\mathbb F_q^n}=\mathcal B(-(I_1+ I_2))\vert_{\mathbb F_q^n}.$ \qed
\end{proof}
\begin{proposition}[BCH Bound] Let $J$ be a defining set of a cyclic code $C$ with minimum distance $d$. If $J$ contains $\delta -1 $ consecutive elements   $\{i,\dots, i+\delta -2\}\subseteq J$, where $i, \delta \in \mathbb{Z}/n \mathbb{Z}$, then $d \geq \delta$. \end{proposition}

The following example illustrates the previous statements on cyclic codes.

\begin{example} Set $q = 2$, $n = 31$.  Let $I_C$ be a generating set and let $J_C$ be a defining set  of the code $C$. The first cyclotomic cosets (modulo 31) are:

$$U_0= \{0\},~ U_1= \{1,2,4,8,16\},~ U_3= \{3,6,12,17,24\}.$$ Consider $C$ the cyclic code with defining set $J_C=U_0 \cup U_1 \cup U_3 $. One has that $J_C$ contains $\{0,1,2,3,4\}$, thus the BCH bound of $C$ is equal to $6$. The dimension of $C$ is equal to $k=|I_C|=31-11=20$. Therefore, the parameters of this code are $[31,20,\geq 6]$.\end{example}

\section{PIR Schemes from Cyclic Codes}\label{sec:principal}

In this section, we focus on cyclic codes towards obtaining PIR schemes over small fields and compute the code parameters with cyclotomic cosets. First, we will analyze the codes obtained from computer search, their cyclotomic cosets, PIR rates, and the number of colluding servers.

The formulation of the amount of colluding servers and the PIR rate is given in  Theorem~\ref{theo:1}. This theorem is   valid for PIR schemes arising from cyclic codes since the automorphism  group of a cyclic code is also transitive~\cite{nine}. %In light of these informations,
Table~\ref{table:I} gives some cyclic codes, the rate of the corresponding PIR scheme arising from them, and the maximum number of servers that may collude, that is, the privacy parameter $t$.

\begin{table}[ht]
\begin{center}
\begin{tabular}{|l|l|l|l|l|l|l|}
\hline
$  C $  & $D$ & $D^{\bot}$ & $C\ast D$ & ${(C\ast D)}^\bot$ & Privacy & Rate  \\ \hline
$   [127,8,63]$ & $[127,29,43]$&$[127,98,10]$ & $[127,113,5]$ & $[127,14,56]$ &9 & $14/127$ \\ \hline
$[127,8,63]$ & $[127,42,32]$&$[127,85,13]$ & $[127,112,6]$ & $[127,15,55]$ &12 & $15/127$ \\ \hline
$[127,15,55]$& $[127,15,55]$ &$[127,112,6]$ & $[127,106,7]$ & $[127,21,48]$ &5 & $21/127$ \\ \hline
$[127,15,55]$ & $[127,21,48]$ &$[127,106,7]$ & $[127,112,6]$ & $[127,15,55]$ &6 & $15/127$ \\ \hline
$[127,21,48]$ &  $[127,21,48]$&$[127,106,7]$ & $[127,112,6]$ & $[127,15,55]$ & 6 & $15/127$  \\ \hline
\end{tabular}
\end{center}
\caption{Computer search experiments}\label{table:I}
\end{table}

For instance, the first row in Table~\ref{table:I} considers $C$ as a storage code  with parameters $[127,8,63]$ and  $D$ as a retrieval code with parameters $[127,29,43]$. Applying  Theorem~\ref{theo:1}, we can conclude that  this scheme is secure against 9-colluding servers since $d(D^\bot)=10$ and that the PIR's rate is $\frac{dim(C\star D)^{\bot}}{n}=\frac{14}{127}$.

We have obtained the codes in Table~\ref{table:I}  by computer search, their generating set can be found in Table~\ref{table:I.A}. For instance, Consider the codes in the first row, the generating set of the code $C$ consists of the union of the cyclotomic cosets $U_1$ and $U_{31}$, and  the one of $D$ consists of $U_0$, $U_5$, $U_{23}$, $U_{27}$, $U_{31}$. As mentioned before, the generating set of star products of cyclic codes are given by the Minkowski sum of their generating sets. Hence, the generating set of $C\star D$ consists of all cyclotomic cosets except $U_{13}$ and $U_{47}$.

From now on, for sake of brevity,  we will denote as $U_{ \{s_1,...,s_t\} }$ the union of the $t$ cyclotomic cosets given by  $U_{ \{s_1,...,s_t\}}$= $\bigcup_{j=1}^{t} U_{s_i}$.
\begin{table}[ht]
\begin{center}
\begin{tabular}{l|l}
\hline
$  C $  &  $D$    \\ \hline
$   U_{\{ 0,31\}}$ & $ U_{\{0,5,23,27,31\}} $   \\
$ U_{\{0,11\}}  $ &  $ U_{\{1,3,11,23,43,55\}}$  \\
$  U_{\{0,5,43\}} $ & $ U_{\{0,23,43\}}$    \\
$  U_{\{0,23,63\}}  $ & $U_{\{19,31,55\}}$    \\
$ U_{\{1,10,29\}}  $ & $U_{\{7,31,55\}}$    \\
\end{tabular}
\end{center}
\caption{Cyclotomic cosets used for   codes in Table \ref{table:I}. }\label{table:I.A}
\end{table}

Table~\ref{table:II} classifies the codes in Table~\ref{table:I} according to the best-known linear codes in the database  \cite{three}, which gives lower and upper bounds on the parameters of linear codes. As it is shown in the following table, their parameters are the best known or optimal.

\begin{table}[ht]
\begin{center}
\begin{tabular}{|l|l|l|l|l|}
\hline
$ C $  & $D$ & $D^{\bot}$ & $C\star D$ & ${(C\star D)}^\bot$   \\ \hline
$optimal$ &  $best-known$ & $best-known$ & $optimal$ & $optimal$ \\ \hline
$optimal$ &  $best-known$ & $best-known$ & $optimal$ & $best-known$ \\ \hline
$best-known$ &  $best-known$ & $optimal$ & $best-known$ & $best-known$ \\ \hline
$best-known$ &  $best-known$ & $best-known$ & $optimal$ & $best-known$ \\ \hline
$best-known$ &  $best-known$ & $best-known$ & $optimal$ & $best-known$ \\ \hline
\end{tabular}
\end{center}
\caption{Classification of codes in Table~\ref{table:I}}\label{table:II}
\end{table}

\subsection{Comparison with Punctured and Shortened RM Codes}

We will show now why cyclic codes may provide better performance than RM codes.
%Reed-Muller codes are linear codes.
Even though a RM code $C$  is not cyclic,  $C_{\bullet}$ and $C^{\bullet}$ are cyclic codes \cite{four}. %and moreover, shortened and punctured RM codes have the same length as cyclic codes.
Therefore, we compare the PIR rate and privacy given by a cyclic code with the corresponding punctured and shortened RM codes.

First, let us focus on the comparison with punctured RM codes. For length $127$ and $255$, we fixed as storage code a $[127, 8, 63]$, $[255, 9, 172]$ cyclic code  and collected the star product of some codes in Table~\ref{table:III} and Table~\ref{table:VI}, respectively. We remark that   in Table~\ref{table:III} the BCH bound of the retrieval codes $(D)$ equal to their minimum distance.% and they have the same parameters of best-known codes in \cite{three}.

\begin{table}[ht]
\begin{center}
\begin{tabular}{|c|c|c|c|c|c|c|}
\hline
$ C $  & $D$ & $D^{\bot}$ & $C\ast D$ & ${(C\ast D)}^\bot$ & Privacy & Rate  \\\hline \rowcolor{Gray}
$[127,8,63]$  & $[127,8,63]$&$[127,119,4]$ & $[127,29,31]$ & $[127,98,7]$ & 3 & $98/127$ \\ \hline
$\textbf{[127,8,63]}$  & $\textbf{[127,22,47]}$&$\textbf{[127,105,8]}$ & $\textbf{[127,64,15]}$ & $\textbf{[127,63,16]}$ & \textbf{7} & $\textbf{63/127}$ \\ \hline \rowcolor{Gray}
$[127,8,63]$  & $[127,29,31]$&$[127,98,8]$ & $[127,64,15]$ & $[127,63,16]$ & 7 & $63/127$ \\ \hline
$\textbf{[127,8,63]}$  & $\textbf{[127,50,27]}$&$\textbf{[127,77,16]}$ & $\textbf{[127,99,7]}$ & $\textbf{[127,28,32]}$ & \textbf{15} & $\textbf{28/127}$\\ \hline
$\textbf{[127,8,63]}$  & $\textbf{[127,57,23]}$&$\textbf{[127,70,16]}$ & $\textbf{[127,99,7]}$ & $\textbf{[127,28,32]}$ & \textbf{15} & $\textbf{28/127}$\\ \hline \rowcolor{Gray}
$[127,8,63]$   & $[127,64,15]$ &$[127,63,16]$ & $[127,99,7]$ & $[127,28,32]$ &15 & $28/127$ \\ \hline
$\textbf{[127,8,63]}$   & $\textbf{[127,85,13]}$ &$\textbf{[127,42,32]}$ & $\textbf{[127,120,3]}$ & $\textbf{[127,7,64]}$ &\textbf{31} & $\textbf{7/127}$ \\ \hline
$\textbf{[127,8,63]}$  & $\textbf{[127,92,11]}$ &$\textbf{[127,35,32]}$ & $\textbf{[127,120,3]}$ & $\textbf{[127,7,64]}$ &\textbf{31} & $\textbf{7/127}$ \\ \hline \rowcolor{Gray}
$[127,8,63]$  & $[127,99,7]$ &$[127,28,32]$ & $[127,120,3]$ & $[127,7,64]$ &31 & $7/127$ \\ \hline
\end{tabular}
\end{center}
\caption{Comparison with punctured RM codes (Shadow rows)}\label{table:III}
\end{table}

\begin{table}[ht]
\begin{center}
\begin{tabular}{l|l}
\hline
$  C $  &  $D$    \\ \hline
$   U_{\{0,1\}}$ & $ U_{\{0,1\}} $   \\
$   $ &  $ U_{\{0,1,5,9\}}$  \\
$  $ & $ U_{\{0,1,5,9,3\}}$    \\
$    $ & $U_{\{0,1,5,9,3,11,19,21\}}$    \\
$   $ & $U_{\{0,1,5,9,3,11,19,21,7\}}$    \\
$   $ & $U_{\{0,1,5,9,3,11,19,21,7,13\}}$ \\
$   $ & $U_{\{0,1,5,9,3,11,19,21,7,13,23,27,43\}}$ \\
$   $ & $U_{\{0,1,5,9,3,11,19,21,7,13,23,27,43,29\}}$ \\
$   $ & $U_{\{0,1,5,9,3,11,19,21,7,13,23,27,29,43,15\}}$ 
\end{tabular}
\end{center}
\caption{Cyclotomic cosets used for   codes in Table \ref{table:III}. }\label{table:III.A}
\end{table}
\begin{table}[ht]
\begin{center} {\renewcommand{\arraystretch}{1.4}
\begin{tabular}{|c|c|c|c|c|c|c|}
\hline
$ C $  & $D$ & $D^{\bot}$ & $C\ast D$ & ${(C\ast D)}^\bot$ & \rotatebox[origin=c]{90}{Privacy} & \rotatebox[origin=c]{90}{Rate}  \\\hline
\rowcolor{Gray}
$[255,9,127]$  & $[255,9,127]$&$[255,246,4]$ & $[255,37,63]$ & $[255,218,8]$ & 3 & $\frac{218}{255}$ \\ \hline
$\textbf{[255,9,127]}$  & $\mathbf{[255,25, \geq  63]}$ & $\mathbf{[255,230, \geq   8]}$ & $\mathbf{[255,93]}$ & $\mathbf{[255,162]}$ & $\mathbf{  \geq  7}$ & $\mathbf{\frac{162}{255}}$ \\ \hline
$\mathbf{[255,9,127]}$  & $\mathbf{[255,33, \geq   63]}$&$\mathbf{[255,222,  \geq  8]}$ & $\mathbf{[255,93]}$ & $\mathbf{[255,162]}$ & $\mathbf{ \geq   7}$ & $\mathbf{\frac{162}{255}}$ \\ \hline
\rowcolor{Gray}
$[255,9,127]$  & $[255,37,63]$&$[255,218,8]$ & $[255,93,31]$ & $[255,162,16]$ & 7 & $\frac{162}{255}$ \\ \hline
$\mathbf{[255,9,127]}$  & $\mathbf{[255,77,\geq     31]}$&$\mathbf{[255,178, \geq   16]}$ & $\mathbf{[255,161]}$ & $\mathbf{[255,94]}$ & $\mathbf{  \geq  15}$ & {$\mathbf{\frac{94}{255}}$}\\ \hline
$\mathbf{[255,9,127]}$  & $\mathbf{[255,85, \geq   31]}$&$\mathbf{[255,170, \geq   16]}$ & $\mathbf{[255,163]}$ & $\mathbf{[255,92]}$ & $\mathbf{  \geq  15}$ & $\mathbf{\frac{92}{255}}$\\ \hline \rowcolor{Gray}
$[255,9,127]$   & $[255,93,31]$ &$[255,162,16]$ & $[255,163,15]$ & $[255,92,32]$ &15 & $\frac{92}{255}$ \\ \hline
$\mathbf{[255,9,127]}$   & $\mathbf{[255,133,  \geq   15]}$ &$\mathbf{[255,122, \geq    32]}$ & $\mathbf{[255,219]}$ & $\mathbf{[255,36]}$ &$\mathbf{  \geq  31}$ & $\mathbf{\frac{36}{255}}$ \\ \hline
$\mathbf{[255,9,127]}$  & $\mathbf{[255,141, \geq   15]}$ &$\mathbf{[255,114, \geq   32]}$ & $\mathbf{[255,219]}$ & $\mathbf{[25536]}$ &$\mathbf{\geq    31}$ & $\mathbf{\frac{36}{255}}$ \\ \hline
$\mathbf{[255,9,127]}$   & $\mathbf{[255,149, \geq   15]}$ &$\mathbf{[255,106, \geq   32]}$ & $\mathbf{[255,219]}$ & $\mathbf{[255,36]}$ &$\mathbf{  \geq  31}$ & $\mathbf{\frac{36}{255}}$ \\ \hline
$\mathbf{[255,9,127]}$  & $\mathbf{[255,153,  \geq  15]}$ &$\mathbf{[255,102,  \geq  32]}$ & $\mathbf{[255,219]}$ & $\mathbf{[25536]}$ & $\mathbf{ \geq   31}$ & $\mathbf{\frac{36}{255}}$ \\ \hline
$\mathbf{[255,9,127]}$  & $\mathbf{[255,161,  \geq  15]}$ &$\mathbf{[255,94,\geq    32]}$ & $\mathbf{[255,219]}$ & $\mathbf{[255,36]}$ & $\mathbf{\geq   31}$ & $\mathbf{\frac{36}{255}}$ \\ \hline \rowcolor{Gray}
$[255,9,127]$  & $[255,163,15]$ &$[255,92,32]$ & $[255,219,7]$ & $[255,36,64]$ &31 & $\frac{36}{255}$ \\ \hline
$\mathbf{[255,9,127]}$  & $\mathbf{[255,211, \geq   7]}$ &$\mathbf{[255,44,  \geq  64]}$ & $\mathbf{[255,247]}$ & $\mathbf{[255,8]}$ &$\mathbf{ \geq   63}$ & $\mathbf{\frac{8}{255}}$ \\ \hline \rowcolor{Gray}
$[255,9,127]$  & $[255,219,7]$ &$[255,36,64]$ & $[255,247,3]$ & $[255,8,128]$ &63 & $\frac{8}{255}$ \\ \hline
\end{tabular}}
\end{center}
\caption{Comparison with punctured RM codes (Shadow rows)}\label{table:VI}
\end{table}
\begin{table}[ht]
\begin{center}
\begin{tabular}{l|l}
\hline
$  C $  &  $D$    \\ \hline
$   U_{\{0,1\}}$ & $ U_{\{0,1\}} $   \\
$   $ &  $ U_{\{0,1,3,5\}}$  \\
$  $ & $ U_{\{0,1,3,5,9\}}$    \\
$    $ & $U_{\{0,1,3,5,9,17\}}$    \\
$   $ & $U_{\{0,1,3,5,9,17,7,11,13,19,25\}}$    \\
$   $ & $U_{\{0,1,3,5,9,17,7,11,13,19,21,25\}}$ \\
$   $ & $U_{\{0,1,3,5,9,17,7,11,13,19,21,25,37\}}$ \\
$   $ & $U_{\{0,1,3,5,9,17,7,11,13,19,21,25,37,15,23,27,29,39\}}$ \\
$   $ & $U_{\{0,1,3,5,9,17,7,11,13,19,21,25,37,15,23,27,29,39,53\}}$ \\
$   $ & $U_{\{0,1,3,5,9,17,7,11,13,19,21,25,37,15,23,27,29,39,45,53\}}$ \\
$   $ & $U_{\{0,1,3,5,9,17,7,11,13,19,21,25,37,15,23,27,29,39,45,51,53\}}$ \\
$   $ & $U_{\{0,1,3,5,9,17,7,11,13,19,21,25,37,15,23,27,29,39,43,45,51,53\}}$ \\
$   $ & $U_{\{0,1,3,5,9,17,7,11,13,19,21,25,37,15,23,27,29,39,43,45,51,53,85\}}$ \\
$   $ & $U_{\{0,1,3,5,9,17,7,11,13,19,21,25,37,15,23,27,29,39,43,45,51,53,85,31,47,55,59,61,87\}}$ \\
$   $ & $U_{\{0,1,3,5,9,17,7,11,13,19,21,25,37,15,23,27,29,39,43,45,51,53,85,31,47,55,59,61,87,91\}}$ 
\end{tabular}
\end{center}
\caption{Cyclotomic cosets used for   codes in Table \ref{table:VI}. }\label{table:VI.A}
\end{table}

Unbold rows in Table~\ref{table:III} and Table~\ref{table:VI}  display the parameters of those codes obtained by the star product of two cyclic codes, equivalent to the punctured RM codes. \textbf{Bold} rows are obtained by the star product of a cyclic code and the fixed code $C$. Consequently, when the rate and the storage codes are fixed, cyclic codes provide the same parameters as punctured RM ones except $D$ with parameters $[255,77,31]$ which gives a better rate than RM codes. However, for a  fixed-length $n$, the dimension of RM codes overgrow, thus for a fixed $C\star D$, there are not many values that the dimensions of the code $C$ and $D$  may take. Hence, the first advantage of using cyclic codes in the PIR scheme is to easily provide a larger constellation of  parameters.

As an  illustration of this fact,  in the fourth and fifth rows of Table~\ref{table:III}, the dimension of $D$ can be  50 or 57  other than 64, or the  dimension of $D$ can be 85 or 92, different than 99. Thus, we have different options for the same rate and privacy.
The following remark will show the method we used for obtaining the codes in Table \ref{table:III} and Table \ref{table:VI}.
\begin{remark} The r-th order punctured generalized RM code is the cyclic code length $n=q^m -1 $ with generator polynomial
\begin{equation} \label{equ:1}
g(x):= \prod_{i \in I} (x-\alpha^i), \: where \; I= \{i: \; w_q(i) \leq (q-1)c \},
\end{equation}
for some  $c \in \mathbb{Z}^+$ and  $w_q(i)$ is the number of non-zeros in the $q$-ary expression of $i$. Now using Equation~(\ref{equ:1}), we have created the unbolded row in Table~\ref{table:III} and Table~\ref{table:VI}. Namely,  if we add or remove some cyclotomic classes to the punctured RM code's generating sets, we can get another cyclic code, which provides the same rate and privacy. % For instance, in the third row, the generating set of $C$ is comprised of $U_0$ and $U_1$, and $D$ is comprised of $U_0$, $U_1$, $U_3$, $U_5$, $U_9$. The generating set of $D$ code in the second row consists of $U_0$, $U_1$, $U_5$, $U_9$ by removing $U_{3}$.
While making these additions and removals of cosets, we use Remark~\ref{rem:1} to decide the heuristics of which  cyclotomic cosets we select. {Note that the minimum distance of the code $D^{\bot}$  provides the privacy of the scheme, so we wish  $d(D^{\bot})-1$ being as big as possible. For this purpose, we set $-I$  to be a large set of consecutive elements. } \\
%\begin{remark}\label{rem:1} Let $S$ be a smallest set of consecutive elements which covers generating set, I, of code $C$ such that $I \subseteq S = \{\mathrm{min}\, I, \mathrm{min}\,  I +1, ..., \mathrm{max}\,  I \} $  where $\mathrm{min}\,  I $ denote the smallest element and $\mathrm{max}\,  I$ denote the biggest element in $I$. Since $\{0, 1, ..., n-1 \} \setminus S$ is the largest set of consecutive elements included in defining set, $J$, minimum distance of $C$ satisfies $d(C) \geq n - |S| +1$, from the definition of BCH bound. This fact is used for in Table~\ref{table:III} and and Table~\ref{table:VI}\end{remark}.
For instance, the third row in Table~\ref{table:III}, the generating set of $C$ is comprised of $U_0$ and $U_1$, and $D$ is comprised of $U_0$, $U_1$, $U_3$, $U_5$, $U_9$. The generating set of code $D$  in the second row consists of $U_0$, $U_1$, $U_5$, $U_9$ by removing $U_{3}$. {We have removed $U_{3}$, because $U_{3}$ does not change the BCH bound of the code. Moreover, in the fourth row in Table~\ref{table:VI}, the generating set of $D$ is comprised of $U_0$, $U_1$, $U_{17}$, $U_9$, $U_5$, $U_3$. Removing $U_{17}$  is not affecting the BCH bound of the code and, thus we obtain the parameter in the third row.}\end{remark}
%Nevertheless, in Table~\ref{table:VI} we use an up to the top procedure, namely ...

We also achieved a second advantage, more privacy, by reducing the dimension of the storage code $C$, which is not equivalent to punctured Reed-Muller codes. We remark that the upload cost in the PIR scheme can be neglected \cite{twelve}, thus we focus on the value $d(D^{\bot})-1$, which provides privacy, and on $dim(C\star D)^{\bot}/n$, which gives the PIR rate. %They are the crucial for the PIR scheme.
Therefore, we can reduce the dimension of the code $C$.

\begin{example}\label{ex:63} Consider the punctured RM codes $C_{RM}$ and $D_{RM}$ with parameters $[63,7,31]$ and $[63,42,7]$, respectively. One has that the product code $C_{RM} \star D_{RM}$ has parameters $[63,57,3]$. The PIR scheme given using $C_{RM}$ and $D_{RM}$ protects against $d_{D^{\bot}}-1 = 15$ collusions. Consider now the cyclic code $C$ with parameters $[63,2,42]$, where the generating set of $C$ is equal to $U_{21}$, and the cyclic code $D$ with parameters $[63,51,3]$. One has that $C \star D  = C_{RM} \star D_{RM}$. In this case $d_{D^{\bot}}-1=19$. Therefore, our cyclic code proposal protects against a more significant number of colluding servers for the same rate.\end{example}

\begin{remark}
Note that for length $63$, there are two good cyclic codes in terms of the PIR parameters, one has    dimension $2$ (see Example~\ref{ex:63}) and the second one, $D$ with parameters $[63,40,7]$, provides the same rate and privacy of a RM code. In the case of length 31, there are no binary cyclic codes that improve the rate or privacy of a RM code other than the ones equivalent to them. This is  why we consider binary cyclic codes  of length greater than or equal to $127$.
\end{remark}
\begin{table}[ht]
\begin{center}
\begin{tabular}{|c|c|c|c|c|c|}
\hline
$  C $  &  $D$ & $D^{\bot}$ & ${(C\star D)}^\bot$ & Privacy & Rate  \\ \hline
$   [255,2,170]$ & $[255,192]$ & $[255,63,\textbf{20+45}]$ & $[255,\textbf{19}]$ & $64$ & $19/255$ \\ \hline
$   [255,2,170]$ &  $[255,195]$ & $[255,60,\textbf{15+51}]$ & $[255,\textbf{8}]$ & $65$ & $8/255$ \\ \hline
$   [255,2,170]$ & $[255,198]$ & $[255,57,\textbf{15+53}]$ & $[255,\textbf{8}]$ & $67$ & $8/255$ \\ \hline
$   [255,2,170]$ &  $[255,200]$ & $[255,55,\textbf{29+41}]$ & $[255,\textbf{11}]$ & $69$ & $11/255$ \\ \hline
$   [255,2,170]$ & $[255,201]$ & $[255,54,\textbf{40+32}]$ & $[255,\textbf{9}]$ & $71$ & $9/255$ \\ \hline
$   [255,2,170]$ &  $[255,202]$ & $[255,53,\textbf{28+44}]$ & $[255,\textbf{8}]$ & $71$ & $8/255$ \\ \hline
$   [255,2,170]$  & $[255,204]$ & $[255,51,\textbf{14+60}]$ & $[255,\textbf{11}]$ & $73$ & $11/255$ \\ \hline
\end{tabular}
\end{center}
\caption{Reducing the dimension of the storage}\label{table:IV}
\end{table}
\begin{table}[ht]
\begin{center}
\begin{tabular}{l|l}
\hline
$  C $  &  $D$    \\ \hline
$   U_{85}$ & $V \setminus (U_{\{0,1,11,13,17,21,25,61,85,87\}}) $   \\
$   $ &  $V \setminus (U_{\{1,13,25,27,29,31,45,119\}})$  \\
$   $ & $V \setminus (U_{\{0,1,7,13,25,31,39,45\}} )$    \\
$   $ & $V \setminus (U_{\{0,1,13,17,25,29,31,63,85\}} )$    \\
$   $ & $V \setminus (U_{\{39,55,61,63,85,87,119,127\}} )$    \\
$   $ & $V \setminus (U_{\{0,1,9,13,25,31,111,119\}} )$    \\
$   $ & $V \setminus (U_{\{0,1,11,13,29,47,85,111\}} )$    \\
\end{tabular}
\end{center}
\caption{Cyclotomic cosets used for   codes in Table \ref{table:IV}, where  $V$ is the set of all cyclotomic cosets for $q=2$, modulo  $n=255$. }\label{table:IV.A}
\end{table}

Table \ref{table:IV} contains more examples where, by using cyclic codes, the dimension of the storage code has been reduced. In this table, { the minimum distance of $D^\perp$, which is related to the privacy, was first evaluated by the BCH bound and then its real value was computed using the powerful minimum distance algorithm in \cite{Fernando} (for instance, by $20+15$ we mean that the BCH bound is equal to $20$ and the real minimum distance is equal to $35$)}.  The table displays the privacy (number of colluding servers) and rate of the PIR scheme obtained using a code $C$ with length $255$ and dimension 2. Note that the PIR scheme obtained using the Punctured RM codes $C_{RM}$ and $D^{\bot}_{RM}$ with parameters $[255,9,127]$ and $[255,36,64]$, respectively, protects against a maximum of $63$ colluding servers. Moreover, the PIR rate of this scheme is equal to $8/255$. In Table \ref{table:IV}, the code pairs in all rows protect against more than 63 collusions. The cyclotomic cosets used for constructing the codes in Table~\ref{table:IV} are given in Table~\ref{table:IV.A}.

%As mentioned before, shortened RM codes are cyclic codes like punctured RM and we will consider them now.
In Table~\ref{table:V}, shortened RM codes at the evaluation of $\mathbf 0$  and cyclic codes with length $127$ are analyzed. Again, we specify \textbf{bold} rows for star product of cyclic codes and unbold rows for star product of shortened RM codes. The storage code $C$ with parameters $[127,7,64]$, equivalent to a shortened RM, is fixed. The only difference with respect to  Table \ref{table:III} is that we do not include the cyclotomic coset $U_{0}$ in the generating set of $C$.

\begin{table}[ht]
\begin{center}{\renewcommand{\arraystretch}{1.4}
\begin{tabular}{|c|c|c|c|c|c|c|}
\hline
$ C $  & $D$ & $D^{\bot}$ & $C\ast D$ & ${(C\ast D)}^\bot$ & \rotatebox[origin=c]{90}{Privacy} & \rotatebox[origin=c]{90}{Rate} \\ \hline \rowcolor{Gray}
$[127,7,64]$  & $[127,7,64]$&$[127,120,3]$ & $[127,28,32]$ & $[127,99,7]$ &2 & $\frac{99}{127}$ \\ \hline
$\mathbf{[127,7,   64]}$  & $\mathbf{[127,22,    47]}$&$\mathbf{[127,105,    8]}$ & $\mathbf{[127,63,    16]}$ & $\mathbf{[127,64,   15]}$ &$  \mathbf{7}$ & $\mathbf{\frac{64}{127}}$ \\ \hline \rowcolor{Gray}
$[127,7,64]$  & $[127,28,32]$&$[127,99,7]$ & $[127,63,16]$ & $[127,64,15]$ &6 & $\frac{64}{127}$ \\ \hline
$\mathbf{[127,7,   64]}$  & $\mathbf{[127,50,   23]}$&$\mathbf{[127,77,   16]}$ & $\mathbf{[127,98,   8]}$ & $\mathbf{[127,29,   31]}$ &$  \mathbf{15}$ & $\mathbf{\frac{29}{127}}$\\ \hline
$\mathbf{[127,7,   64]}$  & $\mathbf{[127,57,   23]}$&$\mathbf{[127,70,   16]}$ & $\mathbf{[127,98,   8]}$ & $\mathbf{[127,29,   31]}$ & $  \mathbf{15}$ & $\mathbf{\frac{29}{127}}$\\ \hline \rowcolor{Gray}
$[127,7,64]$   & $[127,63,16]$ &$[127,64,15]$ & $[127,98,8]$ & $[127,29,31]$ &14 & $\frac{29}{127}$ \\ \hline
$\mathbf{[127,7,   64]}$   & $\mathbf{[127,85,   13]}$ &$\mathbf{[127,42,   32]}$ & $\mathbf{[127,119,   4]}$ & $\mathbf{[127,8,   63]}$ &$   \mathbf{31}$ & $\mathbf{\frac{8}{127}}$ \\ \hline
$\mathbf{[127,7,   64]}$  & $\mathbf{[127,92,   11]}$ &$\mathbf{[127,35,   32]}$ & $\mathbf{[127,119,   4]}$ & $\mathbf{[127,8,   63]}$ &$  \mathbf{31}$ & $\mathbf{\frac{8}{127}}$ \\ \hline \rowcolor{Gray}
$[127,7,64]$  & $[127,98,8]$ &$[127,29,31]$ & $[127,119,4]$ & $[127,8,63]$ &30 & $\frac{8}{127}$ \\ \hline
\end{tabular}}
\end{center}
\caption{Comparison with shortened RM codes (Shadow rows)}\label{table:V}
\end{table}

One has that the PIR schemes using cyclic codes protect against one more colluding server than shortened RM codes, as it can be seen at Table \ref{table:V}. Moreover, in this way we may increase the constellation of possible parameters. For instance, for a case rate equal to $29/127$, the PIR scheme coming from a cyclic code protects against 15-collusion, but the one from a shortened RM code protects against 14-collusion. The cyclotomic cosets used for constructing the codes in Table \ref{table:V} are given in Table \ref{table:V.A}.

\begin{table}[ht]
\begin{center}
\begin{tabular}{l|l}
\hline
$  C $  &  $D$    \\ \hline
$   U_{\{1\}} $ & $ U_{\{1\}} $   \\
$   $ &  $ U_{\{0,1,5,9\}}$  \\
$  $ & $ U_{\{1,5,9,3\}}$    \\
$    $ & $ U_{\{0,1,5,9,3,11,19,21\}} $    \\
$   $ & $ U_{\{0,1,5,9,3,11,19,21,7\}} $    \\
$   $ & $ U_{\{1,5,9,3,11,19,21,7,13\}} $ \\
$   $ & $ U_{\{0,1,5,9,3,11,19,21,7,13,23,27,43\}} $ \\
$   $ & $ U_{\{0,1,5,9,3,11,19,21,7,13,23,27,43,29\}} $ \\
$   $ & $ U_{\{1,5,9,3,11,19,21,7,13,23,27,29,43,15\}} $ 
\end{tabular}
\end{center}
\caption{Cyclotomic cosets used for   codes in Table \ref{table:V}. }\label{table:V.A}
\end{table}

\section{Conclusion}
By using cyclic codes, we provide binary PIR schemes with colluding servers in the fashion of \cite{RM}. We provide a family of optimal binary PIR schemes. Our PIR schemes have the advantage, with respect to PIR schemes from MDS codes, that they can be defined over a binary field. Moreover, they provide a larger constellation of parameters than the binary PIR schemes using Reed-Muller codes and they even outperform them in some cases. Note also that we  come up with a reduced cost in generating the query vectors since a smaller dimension of the retrieval code   means that less randomness will be needed to be generated by the user.  All the examples in the paper were generated using the computer algebra system Magma \cite{Magma}.

\section*{Acknowledgements}
{We would like to thank F. Hernando (Universitat Jaume I) for providing us the code of the algorithm in~\cite{Fernando}}.
%
% ---- Bibliography ----
%
% BibTeX users should specify bibliography style 'splncs04'.
% References will then be sorted and formatted in the correct style.
%
 \bibliographystyle{splncs04}
 \bibliography{PIR.bib}

\begin{thebibliography}{10}
\providecommand{\url}[1]{\texttt{#1}}
\providecommand{\urlprefix}{URL }
\providecommand{\doi}[1]{https://doi.org/#1}

\bibitem{twelve}
Banawan, K., Ulukus, S.: The capacity of private information retrieval from
  coded databases. IEEE Transactions on Information Theory  \textbf{64}(3),
  1945--1956 (2018). \doi{10.1109/TIT.2018.2791994}

\bibitem{Bierbrauer}
Bierbrauer, J.: The theory of cyclic codes and a generalization to additive
  codes. Des. Codes Cryptogr.  \textbf{25}(2),  189--206 (2002).
  \doi{10.1023/A:1013808515797}

\bibitem{Magma}
Bosma, W., Cannon, J., Playoust, C.: The {M}agma algebra system. {I}. {T}he
  user language. J. Symbolic Comput.  \textbf{24}(3-4),  235--265 (1997).
  \doi{10.1006/jsco.1996.0125}, \url{http://dx.doi.org/10.1006/jsco.1996.0125},
  computational algebra and number theory (London, 1993)

\bibitem{Cascudo}
Cascudo, I.: On squares of cyclic codes. IEEE Trans. Inform. Theory
  \textbf{65}(2),  1034--1047 (2019)

\bibitem{origin}
Chor, B., Goldreich, O., Kushilevitz, E., Sudan, M.: Private information
  retrieval. J. ACM  \textbf{45}(6),  965--982 (1998)

\bibitem{RM}
Freij-Hollanti, R., Gnilke, O.W., Hollanti, C., Horlemann-Trautmann, A.L.,
  Karpuk, D., Kubjas, I.: $t$-private information retrieval schemes using
  transitive codes. IEEE Transactions on Information Theory  \textbf{65}(4),
  2107--2118 (2019). \doi{10.1109/TIT.2018.2871050}

\bibitem{siam}
Freij-Hollanti, R., Gnilke, O.W., Hollanti, C., Karpuk, D.A.: Private
  information retrieval from coded databases with colluding servers. SIAM J.
  Appl. Algebra Geom.  \textbf{1}(1),  647--664 (2017)

\bibitem{three}
Grassl, M.: {Bounds on the minimum distance of linear codes and quantum codes}.
  Online available at \url{http://www.codetables.de} (2007), accessed on
  2021-11-03

\bibitem{Fernando}
Hernando, F., Igual, F.D., Quintana-Ort\'{\i}, G.: Algorithm 994: fast
  implementations of the {B}rouwer-{Z}immermann algorithm for the computation
  of the minimum distance of a random linear code. ACM Trans. Math. Software
  \textbf{45}(2),  Art. 23, 28 (2019). \doi{10.1145/3302389},
  \url{https://doi.org/10.1145/3302389}

\bibitem{nine}
Huffman, W.C., Pless, V.: Fundamentals of Error-Correcting Codes. Cambridge
  University Press (2003). \doi{10.1017/CBO9780511807077}

\bibitem{six}
van Lint, J.: Introduction to Coding Theory. Graduate Texts in Mathematics,
  Springer Berlin Heidelberg (1998)

\bibitem{eight}
MacWilliams, F.J., Sloane, N.J.A.: The theory of error-correcting codes. {I}.
  North-Holland Mathematical Library, Vol. 16, North-Holland Publishing Co.,
  Amsterdam-New York-Oxford (1977)

\bibitem{Randriambololona}
Randriambololona, H.: On products and powers of linear codes under
  componentwise multiplication. In: Algorithmic arithmetic, geometry, and
  coding theory, Contemp. Math., vol.~637, pp. 3--78. Amer. Math. Soc.,
  Providence, RI (2015). \doi{10.1090/conm/637/12749},
  \url{https://doi.org/10.1090/conm/637/12749}

\bibitem{five}
Shah, N.B., Rashmi, K.V., Ramchandran, K.: One extra bit of download ensures
  perfectly private information retrieval. In: 2014 IEEE International
  Symposium on Information Theory. pp. 856--860 (2014).
  \doi{10.1109/ISIT.2014.6874954}

\bibitem{SunCapacity1}
Sun, H., Jafar, S.A.: The capacity of private information retrieval. IEEE
  Trans. Inform. Theory  \textbf{63}(7),  4075--4088 (2017)

\bibitem{SunCapacity2}
Sun, H., Jafar, S.A.: The capacity of robust private information retrieval with
  colluding databases. IEEE Trans. Inform. Theory  \textbf{64}(4, part 1),
  2361--2370 (2018)

\bibitem{Tajadine2018}
Tajeddine, R., Gnilke, O.W., el~Rouayheb, S.: Private information retrieval
  from mds coded data in distributed storage systems. IEEE Transactions on
  Information Theory  \textbf{64},  7081--7093 (2018)

\bibitem{four}
Yardi, A.D., Pellikaan, R.: On shortened and punctured cyclic codes. ArXiv
  \textbf{abs/1705.09859} (2017)

\end{thebibliography}

\end{document}